\documentclass[journal=nalefd,manuscript=article]{achemso}

\usepackage[version=3]{mhchem} 

\author{Eugenia Pyurbeeva}
\affiliation {\small \textit School of Physics and Astronomy, Queen Mary University of London, Mile End Road, London E1 4NS, UK}
\author{Chunwei Hsu}
\affiliation {\small \textit Kavli Institute of Nanoscience, Delft University of Technology, Lorentzweg 1, Delft 2628 CJ, The Netherlands}
\author{David Vogel} 
\affiliation {\small \textit Department of Chemistry, University of Basel, St. Johanns‐Ring 19, 4056 Basel, Switzerland}
\author{Christina Wegeberg}
\affiliation{\small \textit Department of Chemistry, University of Basel, St. Johanns‐Ring 19, 4056 Basel, Switzerland}
\author{Marcel Mayor}
\affiliation {\small \textit Department of Chemistry, University of Basel, St. Johanns‐Ring 19, 4056 Basel, Switzerland}
\alsoaffiliation {\small \textit Institute for Nanotechnology (INT), Karlsruhe Institute of Technology (KIT), P.O. Box 3640, 76021 Karlsruhe, Germany}
\alsoaffiliation {\small \textit Lehn Institute of Functional Materials (LIFM), School of Chemistry, Sun Yat-Sen University (SYSU), 510275 Guangzhou, China}
\author{Herre van der Zant}
\affiliation {\small \textit Kavli Institute of Nanoscience, Delft University of Technology, Lorentzweg 1, Delft 2628 CJ, The Netherlands}
\author{Jan A. Mol}
\email{j.mol@qmul.ac.uk}
\affiliation {\small \textit School of Physics and Astronomy, Queen Mary University of London, Mile End Road, London E1 4NS, UK}
\author{Pascal Gehring}
\email{pascal.gehring@uclouvain.be}
\affiliation {\small \textit IMCN/NAPS, Université Catholique de Louvain (UCLouvain), 1348 Louvain-la-Neuve, Belgium}

\title{Controlling the entropy of a single-molecule junction}

\abbreviations{}

\begin{document}



\begin{abstract}
Single molecules are nanoscale thermodynamic systems with few degrees of freedom. Thus, the knowledge of their entropy can reveal the presence of microscopic electron transfer dynamics, that are difficult to observe otherwise.  Here, we apply thermocurrent spectroscopy to directly measure the entropy of a single free radical molecule in a magnetic field. Our results allow us to uncover the presence of a singlet to triplet transition in one of the redox states of the molecule, not detected by conventional charge transport measurements. This highlights the power of thermoelectric measurements which can be used to determine the difference in configurational entropy between the redox states of a nanoscale system involved in conductance without any prior assumptions about its structure or microscopic dynamics. 
\end{abstract}


The thermodynamic concept of entropy is key to the understanding of many chemical processes, including electron transfer reactions \cite{Nitzan2013}. Spontaneous change only occurs when entropy increases and the universe becomes more disordered. While the statistical interpretation of entropy first given by Boltzmann connects entropy as a thermodynamic state function to the number of microstates available to the system, this connection is generally not quantifiable -- macroscopic systems, containing Avogadro's numbers ($\sim10^{23}$) of molecules, simply have too many configurations for the individual to be tractable. At the single-molecule level, however, electron transfer reactions lead to significant changes in configurational entropy related to the electronic structure that can be quantified due to the relatively small number of states of the system \cite{Seifert2012}. The ability to directly measure entropy of individual molecules -- without the need of a priori knowledge of its electronic structure -- can thus provide insights into their non-trivial quantum states, such as (high-)spin ground states \cite{Joeri2019,Coronado2020}, Kondo states \cite{Andrei1984,Hiraoka2017,Guo2021}, Majorana modes \cite{Smirnov2015}, and non-abelian anyons \cite{Ben-Shach2013, Nakamura2020}, and shed light on the discriminating role of entropy in driving electron transfer rates.

Recent experiments have demonstrated the potential for direct entropy measurements in mesoscopic quantum systems \cite{Hartman2018, Sela2019, Kleeorin2019, Gehring2021}. These studies connect macroscopic observable quantities, including electrical current and charge, to microscopic configuration entropy in the form of spin-degeneracy. Here, we scale this approach down to the single-molecule level and determine the redox-dependent entropy of an individual nitronyl nitroxide radical (shown in Figure 1a) directly from its thermoelectric conductance. The entropy measurements indicate that the neutral molecule holds an unpaired electron and is therefore in a doublet ground state, while the ground state of the reduced molecule is a singlet with all electrons paired. However, by applying a magnetic field to the reduced molecule, a low-lying triplet excitation state is revealed which was not observed in conventional charge transport measurements.

\begin{figure}
  \includegraphics{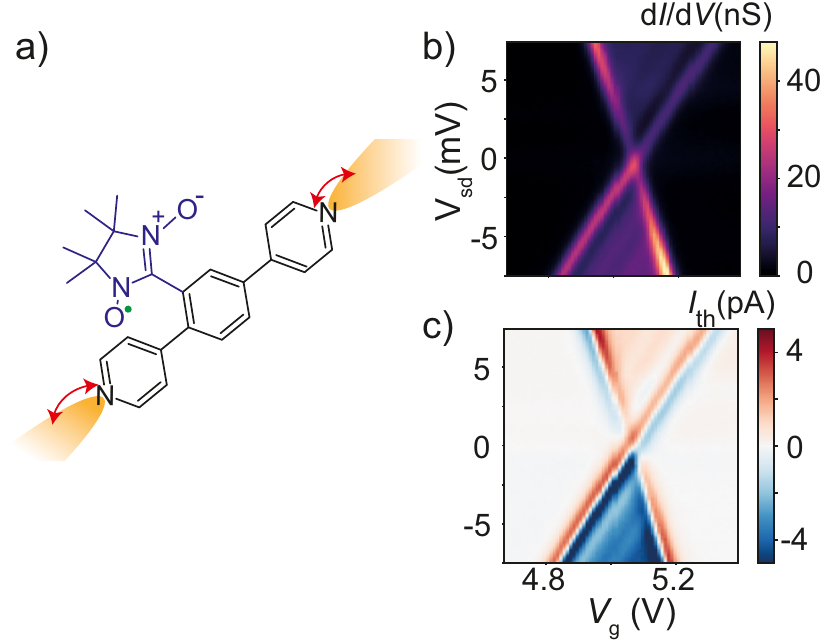}
  \caption{a) Sketch of the molecule (a free radical, green, located at the a nitronyl nitroxide side group, blue, which is connected to a backbone, black) connected to the source and drain lead. The red arrows indicate electron tunneling on and off the molecule. b) Bias and gate voltage dependent differential conductance and c) thermocurrent of a single NNR molecule junction.}
  \label{fig1}
\end{figure}
To measure the entropy difference $\Delta S$ between the neutral (with $N$ electrons) and the reduced (with $N+1$ electrons) free radical molecule directly from its thermoelectric response, we first develop the necessary theoretical framework. To this end, we find the elements of the Onsager matrix that are related to the charge current $I$. Starting from a thermodynamic relation between the internal energy of the molecule, its redox state, and its configurational entropy \cite{Pyurbeeva2021}, we derive an expression for the conductance $G=\partial I/\partial V_{sd}$ (where $V_{sd}$ is the bias voltage across the molecule) from rate equations (see SI) that is consistent with that derived for sequential electron tunneling through a quantum dot\cite{Beenakker1991}. Moreover, we derive an expression for the thermoelectric conductance $L=\partial I/\partial \Delta T$ (where $\Delta T$ is the temperature difference across the molecule) which takes a similar form:

\begin{align}
	\label{fit}
		G \propto \dfrac{1}{T} &\left[1-f(\varepsilon)\right]f(\varepsilon-T \Delta S),\\
	    L\propto\dfrac{\varepsilon}{T^2}&\left[1-f(\varepsilon)\right]f(\varepsilon-T \Delta S),
\end{align}
where $f$ is the Fermi-Dirac distribution, $\varepsilon$ the single-electron energy level with respect to the electrochemical potential of the electrodes, and $T$ the equilibrium temperature of the system. Here we recognise the free energy of formation $\Delta F = \varepsilon-T\Delta S$ associated with the isothermal electrode reaction as the driving force for electron transfer\cite{Marcus1965}. We also observe that, as a result, the conductance is maximum when $\varepsilon = T\Delta S/2$, as was experimentally demonstrated previously\cite{Harzheim2020,Gehring2021}. More importantly, the thermoelectric response is asymmetric around $\varepsilon=0$ and the degree of asymmetry scales directly with $\Delta S$. In the remainder of this Letter, we will therefore exploit this asymmetry as a direct measure for the entropy difference.

The conductance and thermoelectric response of the free radical are measured simultaneously by contacting it in an electromigrated breakjunction with integrated micro-heaters and gate electrodes using a double lock-in technique\cite{Gehring2019, Gehring2021}. Here, we use an organic radical molecule (nitronyl nitroxide radical). It consists of a 1,4-bis(4-pyridyl)benzene backbone and a nitronyl nitroxide side group where an unpaired electron resides as shown in Fig.~\ref{fig1}a. This makes it a single-spin system in its neutral state. Single-molecule junctions are formed by opening a nm-sized gap in a thin gold bridge using electromigration followed by self-breaking\cite{Gehring2019} and  depositing of single-molecules from solution after nano-gap formation. Figure 1c and d show the conductance and thermocurrent $L\Delta T$ as a function of the applied bias ($V_{sd}$) and gate ($V_g$) voltage. We observed regions of low conductance (Coulomb-blocked, black in Figure 1b), corresponding to the $N$ and $N+1$ redox states of the free radical, separated by a region of sequential electron tunneling. Furthermore, lines inside the sequential tunneling regime which run in parallel to the edges of the Coulomb diamond can be observed, which we attribute to vibrational excitations of the molecule \cite{Park2000,Leturcq2009,GehringACS,Gehring2017,Gehring2021}. The thermocurrent (Figure 1c) is zero at the Coulomb diamond edges and reaches minimum/maximum values close to these edges (blue and red lines), similar to our previous findings on Gd-tpy complexes \cite{Gehring2021}.  

\begin{figure}
  \includegraphics[scale=1.0]{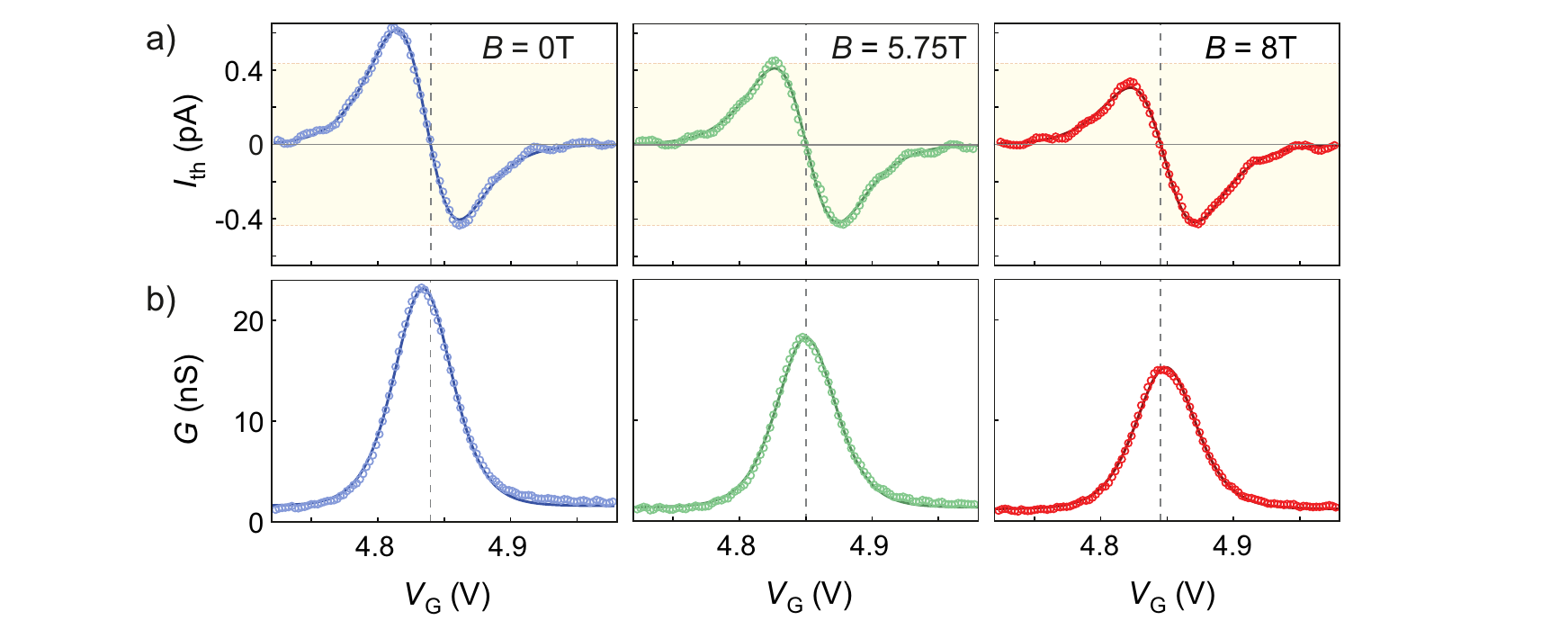}
  \caption{a) Experimental thermocurrent and b) conductance (open circles), and theoretical fits (solid lines) using equations 1 and 2 (in the small temperature difference assumption) at three different magnetic fields. The yellow shaded areas indicate the minimum thermocurrent $I_{\mathrm{th,min}}$. The vertical dotted lines indicate $I_{\mathrm{th}} = 0$.}
  \label{fig2}
\end{figure}

In the following, we will focus on the gate-dependent conductance and thermocurrent traces at zero bias ($V_{sd}=0$), as shown in Figure 2. We find a peak in conductance around a gate value of 4.8 - 4.9V, where an extra charge is continuously added and removed to/from the molecule leading to electron transfer from one side of the molecule to the other. For gate voltages $\gg 4.9$~V $N+1$ charges reside on the molecule and it is in its reduced state. The gate-dependent thermocurrent changes sign around the position of the conductance peak and possesses a pronounced asymmetry between its maximum and minimum. As explained above, this asymmetry is a direct measure of the entropy difference between the redox states of the free radical. We further observe that the degree of asymmetry of the thermocurrent, as well as the shift of the conductance peak, vary when a magnetic field is applied (see Figure 2b and c). This is a clear indication that the entropy of the system can be controlled by a magnetic field. To quantify this entropy change $\Delta S$ as a function of magnetic field we fit the zero-bias conductance and thermocurrent traces shown in Figure 2 to Equations 1 and 2 (see solid lines in Figure 2) using $\varepsilon =\varepsilon_0-\alpha V_g$, where $\alpha$ is the lever-arm given by the capacitance between the free radical and the electrodes \cite{Beenakker1991, Zimbovskaya2013, Gehring2019a}.

\begin{figure}
  \includegraphics[scale=1]{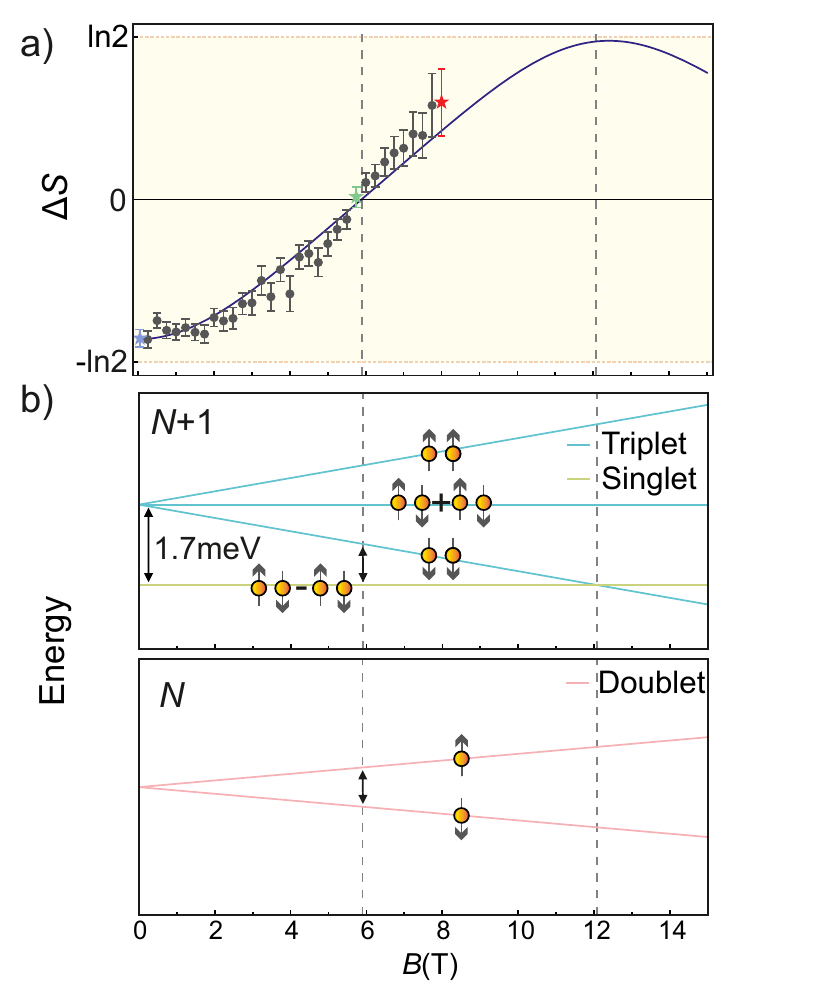}
  \caption{a) Entropy change of the molecule obtained from fitting equation 1 and 2 to the experimental data (in the small temperature difference assumption) as a function of magnetic field (points). Solid line represents the best fit of the entropy change values to the difference in Gibbs entropy between the charge states in the doublet $\rightarrow$ singlet+triplet model. b) Energy level structure for the molecule in its neutral state (N electrons) and reduced state (N+1 electrons) as a function of magnetic field, for the values of $\Delta_{ST}$ and g-factor from fit in \ref{fig3}a. It can be seen that the point of zero entropy change (dashed line at approximately 6T) approximately corresponds to the same energy splitting between the two doublet states and the two lower-lying states of the $N+1$ charge states, meaning that the transition effectively occurs between two identical doublet states (vertical arrows). The yellow spheres (and corresponding arrows) illustrate the spin states of the free radical (on the nitronyl nitroxide unit) and the extra added electron on the backbone (in the reduced case, N+1). }
  \label{fig3}
\end{figure}

Figure 3a shows the resulting entropy difference between the $N$ and $N+1$ redox states as a function of magnetic field determined by fitting to Eqn 2. At zero magnetic field $\Delta S$ approaches a value of $-k_B\ln 2$, where $k_B$ is the Boltzmann constant. Because the Gibbs entropy is equal to $-k_B\sum p \ln p$, where $p$ is the occupation probability of each microstate of the system, we can conclude that the $N$ charge state has 2 accessible microstates with equal occupation probabilities, while the $N+1$ charge state has one. This agrees with the free radical ground state having an unpaired electron in its ground state, and thus being a spin-doublet, and the reduced state having all electrons paired, being a singlet.

When the magnetic field is increased, we observe an increase of $\Delta S$, which reaches a value of zero and changes sign at $B\sim6$~T. With the application of the magnetic field, the degeneracy of the doublet ($N$ electrons) ground state is lifted. As the field -- and with it the Zeeman splitting -- is increased, the probability of occupation of the bottom energy level of the doublet begins to dominate. This can explain the increase of $\Delta S$ at low magnetic fields, however this is not consistent with the subsequent change of sign of $\Delta S$ at approximately 6T. The magnetic-field-dependent entropy therefore points to a low-lying triplet state of the reduced free radical where the entropy associated to the partial population in the $N+1$ singlet and triplet states can exceed the entropy of the $N$ doublet population. We confirm these results by comparing them to previously employed methods \cite{Pyurbeeva2021, Gehring2021} based on the analysis of the difference in the energy levels corresponding to zero thermocurrent point and the conductance peak, where we obtained a similar entropy difference dependence to the thermocurrent asymmetry analysis (see SI section 2).

To determine the singlet-triplet splitting $\Delta_{ST}$ in the $N+1$ redox state, we find the expected entropy difference for the doublet $\rightarrow$ singlet+triplet transition, assuming a Gibbs distribution in each of the charge states, as a function of magnetic field, singlet-triplet splitting $\Delta_{ST}$ and the g-factor. We use this dependence to fit the experimentally obtained $\Delta S$ dependence on magnetic field with $\Delta_{ST}$ and the g-factor as a fitting parameters (the fitting result is shown in Figure 3). This approach yields a $g$-factor of 2.7, deviating from the value of 2 for a free electron. Such deviations have been found before in nano-scale junctions and have been attributed to orbital contributions.\cite{Liu15,Winkler17} Furthermore, we extract a singlet-triplet splitting $\Delta_{ST}=1.7\pm0.1$~meV. This low energy can be explained by assuming that the extra electron (when reducing the molecule) is added to the molecular back-bone rather than to the nitronyl nitroxide side group of the molecule. $\Delta_{ST}$ would then correspond to the exchange interaction between the two free electrons (on the side group and the backbone), similar to what has been found experimentally in similar free radical all-organic systems \cite{Gaudenzi2017}.

The small singlet-triplet splitting, and the resulting entropy contribution of the non-zero $N+1$ triplet population, is the reason for the deviation of $\Delta S$ from $-k_B\ln2$ at zero magnetic field. Moreover, at $B\sim6$~T the splitting between the $\vert T_+\rangle$ triplet state and the singlet is equal to the  splitting between the doublet states (see Figure 3b), and the Gibbs entropy of the two redox states are approximately equal, $-k_B\left[p_{\uparrow}\ln p_{\uparrow} + p_{\downarrow}\ln p_{\downarrow}\right]\approx -k_B\left[p_{S}\ln p_{S} + p_{T_{+}}\ln p_{T_{+}}\right]$, (where $p$s are the occupation probablities of $\uparrow$ -- the spin up doublet state, $\downarrow$ -- spin down doublet state, $S$ -- the singlet state and $T_{+}$ -- the lowest-lying triplet state), as the population in $\vert T_0\rangle$ and $\vert T_-\rangle$ are negligible, where $T_0$ and $T_-$ are the middle- and high-energy triplet states respectively. When $\Delta S=0$, we observe an anti-symmetric thermoelectric response ($I_{\mathrm{th}}(\varepsilon)=-I_{\mathrm{th}}(-\varepsilon)$) as there is no difference in free energy between the two redox states. We expect $\Delta S$ to reach its positive maximum at $B\sim12$~T, outside of our experimental range, when the $\vert T_+\rangle$ and $\vert S\rangle$ states cross and the  $N+1$ entropy is maximum $S_{N+1}\approx k_B\ln 2$ while $S_N$ tends to zero.

This analysis demonstrates that thermocurrent spectroscopy is a sensitive tool to access the fundamental thermodynamic properties of nanoscale devices. It has several advantages over previously employed methods, because, unlike e.g. the approach taken in \cite{Hartman2018, Sela2019} our method does not rely on charge state measurements and therefore does not require the fabrication of complicated charge sensors and is applicable to any devices on which thermoelectric measurements can be performed. Furthermore, it does not require measurements of the device at multiple temperatures to find the entropy, as was the case one of the previous approaches \cite{Gehring2021}. We applied the method to a single-molecule device and found that the direct measurement of the entropy difference with the change of the redox state  reveals the presence of a low-lying triplet state in the molecule and allows us to quantify the exchange coupling between the free radical and the extra added electron. This serves as further evidence for the potential of practical applications of entropy measurement methods.

\begin{acknowledgement}
We thank M. van der Star for his support in sample fabrication. This work is part of the Organization for Scientific Research (NWO) and the Ministry of Education, Culture, and Science (OCW). P.G. (research associate) acknowledges financial support from the F.R.S.-FNRS of Belgium and a Marie Skłodowska-Curie Individual Fellowship under Grant TherSpinMol (ID: 748642) from the European Union’s Horizon 2020 research and innovation programme. J.A.M. was supported through the UKRI Future Leaders Fellowship, Grant No. MR/S032541/1, with in-kind support from the Royal Academy of Engineering. H.v.d.Z, C.H., M.M. and D.V. acknowledge funding by the EU (FET-767187-QuIET). M.M. acknowledges support from the Swiss National Science Foundation (SNF grant numbers 200020-178808) and the 111 project (90002-18011002). C.W. thanks the Independent Research Fund Denmark for an international postdoctoral grant (9059-00003B). 
\end{acknowledgement}

\begin{suppinfo}

This will usually read something like: ``Experimental procedures and
characterization data for all new compounds. The class will
automatically add a sentence pointing to the information on-line:

\end{suppinfo}

\bibliography{refs}

\end{document}